\documentstyle[12pt,aasms4,flushrt]{article}

\tighten

\def\d{{\rm d}}

\input psfig

\lefthead{Allende Prieto, Ruiz Cobo, and Garc\'{\i}a L\'opez}
\righthead{Late-Type Stars Model Photospheres}
\begin{document}

\title{Model Photospheres for Late-Type Stars from the Inversion of 
High-Resolution Spectroscopic Observations. The Sun}

\author{Carlos Allende Prieto, Basilio Ruiz Cobo, and Ram\'on J. Garc\'{\i}a
L\'opez}

\affil{Instituto de Astrof\'\i sica de Canarias, \\ E-38200 La Laguna,
Tenerife,  \\  Spain}

\authoraddr{IAC, E-38200 La Laguna, Tenerife,  SPAIN}

\authoremail{callende@iac.es, brc@iac.es, and rgl@iac.es}

\begin{abstract}

An inversion technique has been developed to recover LTE
one-dimensional model photospheres for late-type stars from very high
resolution, high signal-to-noise stellar line profiles. It is
successfully applied to the Sun by using a set of clean \ion{Ti}{1},
\ion{Ca}{1}, \ion{Cr}{1}, and \ion{Fe}{1} normalized line profiles with
accurate transition probabilities and taking advantage of the well
understood collisional enhancement of the wings of the \ion{Ca}{1} line
at 6162 \AA.  Line and continuum center-to-limb variations, continuum
flux, and wings of strong metal lines are synthesized by means of the
model obtained and are compared with solar observations, as well as
with predictions from other well known theoretical  and empirical solar
models, showing the reliability of the inversion procedure. The
prospects for and limitations of the application of this method to
other late-type stars are discussed.

\end{abstract}
 

\section{Introduction} 
\label{sec1}

Model photospheres for late-type stars are a fundamental cornerstone
of modern Astronomy. The Sun, our closest star, belongs to this stellar
category and its atmosphere provides an exceptional plasma laboratory
to study detailed physical processes. Late-type stars are also the most
numerous group in the Galaxy. They not only dominate its present
dynamics, but their photospheric abundances constitute a unique tool for
tracing its chemical evolution history. Furthermore, the oldest unevolved
stars in the Galaxy, which provide information about the physical
conditions at early epochs and constraints on models of primordial
nucleosynthesis, are also late-type stars.

In this century, considerable efforts have been devoted to theoretical
and empirical modeling of late-type stellar atmospheres. In this
respect, particular success was achieved in the seventies, and most of
the models used nowadays are mainly based on work carried out in that
decade. For a comprehensive review of the state of the art in the
modeling of photospheres of F- and later-type stars see, for example,
the discussion by Gustafsson \& J\o rgensen (1994) and Avrett (1996).
Classical theoretical models are usually computed under the simplifying
assumptions of hydrostatic equilibrium, plane-parallel stratification,
local thermodynamic equilibrium (LTE), and conservation of energy flux
(treating convection  with the mixing length approximation).
High-resolution, high signal-to-noise ratio spectra over extensive
spectral ranges can be used to test the validity of existing
theoretical models and to better constrain the observational features
which have to be reproduced (line profiles and asymmetries, excitation
and ionization equilibria of different elements, stellar fluxes,
etc.).

These data can also be used to construct models by using observed
features to find a temperature--depth ($T-\tau$) relation.
Semi-empirical models of this kind have been obtained for the solar
photosphere, where the present-day knowledge of the abundances of
different elements, as well as its extended disk, allow a better
observational constraint. The model constructed by Holweger \& M\"uller
(1974) is a good example of this class and has been widely used not
only for the Sun but also for other late-type stars by scaling its
$T-\tau$ relation to different stellar effective temperatures. Although
the spatially resolved information that can be obtained for the Sun is
lacking for other late-type stars, spectral lines do contain enough
information to determine reliable values for the physical magnitudes
governing the state of a stellar atmosphere, and this approach has been
applied, for instance, to the modelling of giant stars such as Arcturus
(M\"ackle et al. 1975) and Pollux (Ruland et al. 1980).

The so-called spectroscopic {\it inversion} methods are aimed at
obtaining the model atmosphere from which the synthetic line profiles
best match the observed ones (usually with a least-squares criterion).
The availability of high-quality data (observed spectra and line
parameters) and modern computers make possible  these inversions by
taking into account the information contained in the whole line
profile, which can potentially provide not only the temperature
stratification in the atmosphere but also insights into other physical
magnitudes such as velocity and magnetic fields.

This paper is part of a wider project aimed at better understanding the
photospheres of late-type stars of different metallicities for use in
studies of Galactic chemical evolution and primordial nucleosynthesis.
Among the different approaches envisaged within this project, which
incorporates the study of theoretical model atmospheres and the
presence and extension of line asymmetries in the photospheres of
metal-poor stars, we present here a successful attempt to obtain a
semi-empirical solar model photosphere by inverting observed clean line
profiles with accurate oscillator strengths. This kind of information
is also available for other late-type stars (with different levels of
metallicity), and what we learn from modeling the Sun will help us when
trying to apply the method to other stars.  Section 2 provides the
details of the inversion code, which is based on a well-tested
inversion method for the Stokes parameter vector; \S 3 shows its
application to the solar case and the comparison of its results with
those associated with other well-known theoretical and semi-empirical
models. We describe in \S 4 the potential use and limitations in
applying the code to other late-type stars, and the main conclusions
and the perspectives for the near future are summarized in \S 5.

\section{The inversion code}
\label{invcode}

The MISS (Multiline Inversion of Stellar Spectra) code described in
this paper is an adaptation of a previous inversion method of  Stokes
vector spectra:  the SIR method (Ruiz Cobo \& del Toro Iniesta 1992;
see also del Toro Iniesta \& Ruiz Cobo 1995, 1996, 1997). The SIR code
has recently been applied to the study of different solar structures
observed in polarized light, such as sunspots (Collados et al. 1994;
Westendorp Plaza et al. 1997a,b,c), and unresolved magnetic elements in
facular regions (Bellot Rubio, Ruiz Cobo \& Collados 1997), and for
non-polarized light to structures such as penumbrae (del Toro Iniesta,
Tarbell \& Ruiz Cobo 1994), solar granulation (Ruiz Cobo et al. 1995,
1996; Rodr\'\i guez Hidalgo, Ruiz Cobo \& Collados 1996), and the solar
5-min oscillation (Rodr\'\i guez Hidalgo, Ruiz Cobo \& Collados 1997;
Ruiz Cobo, Rodr\'\i guez Hidalgo \& Collados 1997). A study of its
performance versus other inversion techniques can be found in Wood \&
Fox (1995) and Westendorp Plaza et al. (1998).

Our inversion method iteratively modifies an initial model atmosphere
until an optimal fit of the observed profiles is reached. The model
atmosphere contains a set of values of temperature,  microturbulence,
and line-of-sight velocity  evaluated on an equally spaced logarithmic
scale of the optical depth of the continuum at 5000 \AA\ ($\tau_{5000}$), 
together with a set of optical depth-independent parameters such
as rotation velocity ($v\sin \hat{\i}$), macroturbulent velocity
amplitude, and chemical abundances.  Gravity is fixed and the gas and
electronic pressures are computed to satisfy hydrostatic equilibrium,
the ideal gas equation of state, and the Saha equation.

In this section we will pay special attention to
the differential features between our inversion code (MISS) and the
progenitor (SIR), describing its particularities  within three
subsections dedicated to hypotheses and model parameterization,
spectral synthesis, and inversion algorithm.

\subsection{Hypotheses and model parameterization}
\label{hypotheses}

In order to simplify the synthesis process, as well as to keep the number of
free parameters small enough, a set of hypotheses has been assumed:
the LTE approximation, a steady-state, plane-parallel, one-dimensional
atmosphere, solid body rotation, hydrostatic equilibrium, radial
bulk velocity, and negligible magnetic field.

To evaluate the continuum absorption coefficient, contributions  from H,
He, H$^-$, He$^-$, H$_2^-$, H$_2^+$, C, Mg, Na, and scattering terms
have been taken into account. Other atomic and molecular line opacity
sources, which are neglected here, can also be included.

To obtain an accurate solution of the transfer equation the number of
points (optical depth values) through the atmosphere has to be about
50.  Considering the values for every one of the depth-dependent
quantities  as free parameters would imply solving a system of more
than 150 dimensions.  To simplify the numerical problem, following the
SIR strategy, the perturbations of depth-dependent quantities are
evaluated only in a discrete number of points in $\tau_{5000}$, called
{\it nodes}. The entire stratifications of the various parameters are
thus obtained by adding the result of a spline interpolation of the
perturbations at the nodes to the initial stratification. For each
parameter a different number of  nodes can be selected by the user.
Experience from the quoted applications of SIR reveals that the optimum
solution is reached by steadily increasing the degree of complexity of
perturbations (i.e., the number of nodes) from one iteration to the
next, until no significant improvement in the fit between the observed
and synthesized profiles is reached.

Provided the quantity and quality of observational data allow a
corresponding increase in the number of free parameters, some of the
previous hypotheses can be relaxed in the framework of the present MISS
code: the homogeneous atmosphere can be substituted for a two-component
atmosphere in order to take the effect of convective motions into
account; the hydrostatic-equilibrium hypothesis can be ignored by
considering the electron-pressure stratification as a free parameter;
the application of MISS to the determination of magnetic fields is
straightforward, since this code is a simplified version of SIR, a
procedure for Stokes vector inversion. Nevertheless, as will be shown
below, the amount of information we shall be able to extract is
strongly limited by problems deriving from the accuracy of atomic
parameters (mainly oscillator strengths), as well as from the presence
of observational noise. Therefore, up to the present time, the search
for inhomogeneous, magnetized or far from hydrostatic equilibrium
models is beyond our reach. Concerning the LTE approach, a project is
at present in progress for inversion of the Stokes parameter vector
under the following hypotheses: a steady-state atmosphere, complete
redistribution on scattering, LS-coupling, and no quantum interference
between Zeeman states (Socas-Navarro et al. 1996, Socas-Navarro, Ruiz
Cobo \& Trujillo Bueno 1998).

\subsection{Spectral synthesis}
\label{synthesis}

A typical application of MISS to spectroscopic observations implies the
solution of the radiative transfer equation a large number of times.
There are several reasons for this:  the inversion problem must be
solved iteratively due to the non-linear dependence of the spectral
profiles on the physical quantities; to evaluate the stellar flux, the
solution of the radiation transfer equation must be calculated for
several lines of sight; and finally many spectral lines are needed to
obtain a reliable model. Therefore, spectral synthesis is the most
time-consuming part of the code, and it is of paramount importance to
have a fast integration method. MISS uses a new one, recently developed
by Bellot Rubio, Ruiz Cobo \& Collados (1998), which is based on the
Taylor expansion of the Stokes parameter vector (only the specific
intensity in our case) to fourth order in depth. Compared to other
methods, the new technique turns out to be superior in terms of speed
and accuracy. It also gives an approximation to  response functions
(RFs, the derivatives of specific intensity with respect to the
perturbation of each free parameter; see Ruiz Cobo \& del Toro Iniesta
1992, 1994), which are a fundamental ingredient of the inversion
algorithm.

The following quadrature has been used to evaluate the energy flux,
$F$, at a given wavelength, $\lambda$, from the specific intensity,
$I$, computed at each point $\mu$ across the stellar disk:

\begin{equation}
\label{cuadr}
F(\lambda)=\int_0^1 \mu \;I(\lambda,\mu)\; \d \mu \approx \sum_i^n w_i \;
I(\lambda,\mu_i),
\end{equation} 

\noindent where $\mu_i$ are the abscissae and $w_i$ are the weighting
factors for Gaussian integration. Using solar model atmospheres from
the literature, it has been numerically checked that a quadrature of order
$n\approx5$  is enough to guarantee that the integration precision
exceeds the observational data accuracy.

When taking into account the influence of the instrumental profile,
$P(\lambda)$, macroturbulence, $M(\lambda)$, and rotation,
$G_i(\lambda)$, for the i-th annulus (see Appendix A for details),
equation (\ref{cuadr}) becomes:

\begin{equation}
\label{flujo}
F(\lambda)= \left[ \sum_i^n w_i \; I(\lambda,\mu_i) \ast G_i(\lambda) \right]
\ast P(\lambda) \ast M(\lambda),
\end{equation} 

\noindent where ``$\ast$'' means convolution. The instrumental profile,
including telescope and spectrograph, can be provided either as input or
approximated by a Gaussian in which $\sigma$ depends on the spectral
resolution. The macroturbulence profile is a Gaussian with $\sigma$
proportional to the amplitude of the macroturbulent velocity.

\subsection{Inversion algorithm}
\label{algor}

The algorithm is the same as used by SIR, and a detailed description of
it can be found in the above mentioned references. A brief overview is
presented below.

Inversion of observed data is performed by using a non-linear
least-squares algorithm which involves the minimization of a merit
function:

\begin{equation}
\label{chi2}
\chi^2\equiv\frac{1}{\nu}\sum_{i=1}^{i=M} [F^{\rm obs}(\lambda_i)-F^{\rm
syn}(\lambda_i)]^2,
\end{equation}

\noindent where  index $i =1,2, \ldots,M$ stands for the wavelength
samples, indices ``obs'' and ``syn'' refer  to observed and
synthetic data, respectively, and $\nu$ is the number of degrees of
freedom.

Minimization of the merit function is carried out through a Marquardt
algorithm (Marquardt 1963), modified by applying a singular value
decomposition (SVD; Press et al 1986). This algorithm requires the
evaluation of the derivatives of $\chi^2$ with respect to each free
parameter. These derivatives can be directly expressed in terms of 
response functions, which are calculated by applying a
quadrature (eq. [\ref{cuadr}]) to the RF of the specific intensity provided
by the integration method.  Elimination of singularities (via SVD in
this case) is essential because certain free parameters (e.g.,
temperature in layers where the line profile has no sensitivity) may
produce a null effect on the emergent flux and consequently make the
problem undetermined.

To summarize, the steps of the procedure are:

\hspace{3pt} 1. Choice of a guess model and a number of nodes for each free
parameter.

\hspace{3pt} 2. Calculation of flux, RF, and $\chi^2$.

\hspace{3pt} 3. Iterative evaluation of a new model after application of the
Marquardt algorithm, until minimization of $\chi^2$ is reached.

\hspace{3pt} 4. Subsequent modification of the number of nodes and
iteration from step 2 until  $\chi^2$ ceases to decrease
significantly.

\section{Application of the method: The Sun}
\label{sec3}

The SIR method has been successfully applied to the Sun taking
advantage of its spatially resolved disk and the high quality of solar
observations (see references in \S 2). The Sun is also the best known
late-type star and provides a unique opportunity to  field-test the
potential of the MISS technique for the Sun seen as a star.

\subsection{Input data}

The widely used high-resolution Solar Flux Atlas of
\markcite{kuruczatlas84}Kurucz et al. (1984) was selected as the source
of the line profiles. Its complete spectral coverage in the
2960--130000 \AA\ region, with a very high signal-to-noise ratio
($\sim$ 2500), and the extremely high spectral resolution
(${\lambda}/{\Delta\lambda}\sim$ 400000) makes it ideal for our
purposes.  Furthermore, Allende Prieto \& Garc\'\i a L\'opez (1998)
have recently verified the reliability of the wavelength calibration of
this atlas by measuring the central wavelength shifts of 1446
\ion{Fe}{1} lines with accurate laboratory determinations.
        
It was required that the input lines to be included in the inversion be
selected from the  compilation of solar lines by
\markcite{meylan}Meylan et al.  (1993), who identified clean line wings
in the same atlas by fitting Voigt profiles. We also impose the
condition  that the atomic transition probability had been measured at
the Oxford furnace (see, e.g., \markcite{blackwell}Blackwell \& Shallis
1979). Their oscillator strengths ($\log \,gf$) are usually recognized
as the most accurate ones present in the literature. Several tests were
carried out to study the dependence of the method on the accuracy of
the oscillator strengths used. It turned out that when artificially
introducing errors in the $\log gf$ values larger than $\sim$ 5 \% the
inversion returns errors larger than 2\% in the temperature
stratification.  Furthermore, special attention was given to the
chemical elements selected.  We mainly focused on lines of elements
which show an agreement between their photospheric and meteoritic
abundances, as it is more likely for the abundance errors to be
smaller. Although iron does not satisfy this condition, it was felt
that the significant increase in the number of useful lines justified
its inclusion. Finally, in an effort to minimize the problems related
to the lack of knowledge of the collisional damping of the lines,
features stronger that 80 m\AA\ (equivalent width) were rejected.

The method is able to extract information only on the photospheric
region corresponding to the depth range where the employed lines are
formed. To improve the sampling of the photospheric domain, whole line
profiles or just single wings were accepted as input data following the
considerations of the line cleanness by \markcite{meylan}Meylan et al.
(1993).

A total of 40 absorption lines produced by neutral iron, titanium,
chromium and calcium entered the inversion program. The wings of the
stronger \ion{Ca}{1} $\lambda$ 6162 \AA\ were included with the
aim of extending the depth coverage. This is one of the few spectral
lines for which the collisional enhancement of the wings is well known
(see \markcite{pressure}Allende Prieto, Garc\'{\i}a L\'opez \& Trujillo
Bueno 1997), there being a reasonable agreement between theoretical
analyses and semi-empirical measurements.  This consideration is
mandatory in order to  wide the input information without introducing
new uncertainties. The collisional width of the line was interpreted in
terms of collisions with neutral hydrogen, as computed by Spielfiedel
et al. (1991). Table 1 lists the lines entering the inversion and their
parameters.

An attempt was made to employ the spectral information contained in both
the continuum and the line shapes. The observed continuum may differ
strongly from the true one in the blue ($\lambda <$ 5000 \AA) due to
the line absorption, so we can determine only a lower limit for the flux
at these wavelengths. Even the measurement of the pseudo-continuum is a
non-trivial matter: the discrepancies among different authors are as
high as 8\% (see, e.g.,  \markcite{burlov}Burlov-Vasiljev, Gurtovenko \&
Matvejev 1995).  The slope of the continuum is closely linked to the
temperature gradient in the photosphere in such a way that the
employment of the information stored in the continuum requires highly
accurate spectrophotometry. This led us to exclude the absolute
calibration of the line profiles entering the inversion code, and only
line profiles normalized to their local continuum were used.

\subsection{Procedure}

Starting from an isothermal model photosphere, step by step the depth
dependence of the temperature is allowed to be modified in a
successively increasing number of nodes. No vertical velocity gradient
was allowed for, and the microturbulence was assumed to be constant
with depth. A Gaussian profile was employed to represent the
instrumental profile. Three snapshots of the inversion process are
shown in Fig. 1 (ordered a, b, and c). The evolution of the synthetic
profiles for several of the lines included in the inversion is shown
and compared with the solar observations (dots). A constant-temperature
model photosphere (dotted line) does not produce any line absorption,
while the final model (solid line) provides lines which get very close
to the observed features. Starting from the solar abundances listed by
\markcite{ange}Anders \& Grevesse (1989),  the successive modification
of the abundance of the considered elements was allowed for, providing
an excellent final fit. This resulted in no variation of any of the
abundances  except for iron, which dropped from the initial value log
N(Fe) $=$ 7.67 to 7.48 (where log N(H) $=$ 12), very close to the
meteoritic abundance.  Using the synodic equatorial velocity of 1.88 km
s$^{-1}$ and the microturbulence  of 0.6 km s$^{-1}$ (consistent with
Takeda 1995), the code arrives at 1.7 km s$^{-1}$ for the
macroturbulent velocity.

As an example, one of the lines employed in the inversion procedure is
displayed in more detail in Fig. 2, comparing the final synthetic
profile (solid line) with the solar feature (circles).  The point has
been
 reached where no closer fit is possible without taking the line
 asymmetries into account. The observed asymmetries have their origin
mainly in the correlation between temperature and convective velocities,
as observed in the solar granulation, although blends, isotopic or
hyperfine splitting might be responsible for particular cases (Gray
1980; Dravins, Lindegren \& Nordlund 1981; Kurucz 1993). A more complex
treatment than a plane-parallel one-dimensional static single-component
model photosphere is required to reproduce it (e.g., Dravins \&
Nordlund 1990a,b; Rast et al. 1993).

Figure 3 shows the final model photosphere, which we will refer to as
MISS hereafter, compared with other well-known models from the
literature: Kurucz (1992), Holweger \&  M\"uller (1974; hereafter
HOLMU), Vernazza, Avrett \& Loeser (1981; hereafter VALC), and
Gingerich et al. (1971; Harvard Smithsonian Reference Atmosphere,
hereafter HSRA). Kurucz's model is based on theoretical calculations
assuming LTE, hydrostatic equilibrium and taking the line
blanketing into account.  The model of Howeger \& M\"uller
was empirically adjusted to reproduce continuum  observations in the
optical and near infrared, as well as equivalent widths and residual
intensities of \ion{Fe}{1} spectral lines.  The VALC empirical model
was based on observations of the ultraviolet continuum, taking into
account the departures from LTE. And HSRA was basically designed to
reproduce observations of the solar continuum.

The error bars shown for the MISS model have been computed following
the recipe described in Ruiz Cobo et al. (1997) and  limit the depth
range where the lines employed sample the atmosphere; that is, where
the model is meaningful. The MISS model shows, in general, a very
similar temperature dependence to that of the other models, tending to
be cooler in the upper part of the photosphere and hotter in the inner
part.

\subsection{Properties of the model derived}

Insight into the quality of the model photosphere obtained can be
achieved by comparing the behavior of predicted and observed features
not included in the empirical modeling. Choosing what we consider as
the most significant features, systematic comparisons with other
well-established model photospheres for the Sun are shown below.

\subsubsection{Limb darkening}

The center-to-limb variation observed in the Sun maps the temperature
stratification of the solar photosphere and hence is probably one of
the best tools for testing a one-dimensional model photosphere. Figure
4 shows the comparison at $\lambda\lambda$ 3500, 4163, 4774, 5799,
7487, and 10990 \AA\ between polynomial fits to the continuum
observations of \markcite{neckel94}Neckel \& Labs (1994) and the
predictions from the different models considered. The difference
between the observed and computed intensities against $\mu$ (the cosine
of the position angle on the solar surface with respect to our line of
sight) is plotted, normalized to their corresponding values at the disk
center.  HOLMU reproduces in detail the observed variation at $\sim 1$
$\mu$m, the agreement being worse at shorter wavelengths. To make a fully legitimate comparison at short optical wavelengths ($\lambda <$
5000 \AA), it will be necessary to clarify first whether or not the
thousands of lines in the blue part of the optical spectrum do
seriously change the observed limb-darkening from its shape in the true
continuum. HSRA provides the best general comparison at all
wavelengths, while the limb-darkening predicted by the MISS model keeps
close to the observations for the whole optical range.

There are observations available for the variation of line profiles
across the disk. \markcite{balthasar88}Balthasar (1988) lists the
center-to-limb variation for the equivalent widths of 143 lines, and
shows that the behavior of the different lines is not homogeneous. A
comparison between the observations and predictions from the model
atmospheres is shown in Figure 5. The variation of the equivalent
widths normalized to the center of the disk is shown for three lines in
common between our line list and Balthasar's measurements. It can be
seen how the absolute synthetic equivalent widths (with values in the
range 70--100 m\AA) are underestimated with respect to the observed
ones, arising from the use of the Van der Waals approach for the
collisional enhancement of the line wings. Although the chemical
abundances of the different elements are able to alter the curves, and
the temperature dependence of the collisional width is more likely to
be T$^{0.4}$, rather than the Van der Waals dependence of T$^{0.3}$
(e.g., \markcite{pressure}Allende Prieto et al. 1997), we deduce from
the figure that the model obtained from the inversion behaves
remarkably better than the others. Caution must be taken regarding the
HSRA and VALC models: both were constructed under NLTE and the
synthetic profiles shown here were calculated under LTE. A more
precise way of carrying out this comparison would be to use directly
the wing profiles of lines with accurate broadening parameters, such as
the sodium D lines and the \ion{Ca}{1} $4s4p$--$4s5s$ lines at
$\lambda\lambda$ 6102.72, 6122.22, and 6162.17 \AA.

\subsubsection{Continuum flux}

As previously noted, although there is some disagreement in the
literature, the absolute solar flux can be used as a lower limit for
the true continuum. This may help to discard model atmospheres
predicting lower true continuum fluxes than the measured
pseudo-continuum. Figure 6 shows the comparison between the
measurements by \markcite{neckel84}Neckel \& Labs (1984; triangles),
the pseudo-continuum estimated from the correction of the line
absorption observed in the high-resolution solar spectrum of
\markcite{kuruczatlas84}Kurucz et al. (1984; dots), and the theoretical
true continua as predicted by several classical solar models.
\markcite{kurucz92}Kurucz (1992) claims that his solar model reproduces
the solar observations of \markcite{neckel84}Neckel \& Labs (1984) with
high accuracy and, according to this, the true continuum calculated by
taking no line absorption into account  is well over the
pseudo-continuum. That is not the case for the continua associated with
two of the empirical models, HSRA and VALC, being the continuum
predicted by HOLMU higher than the pseudo-continuum but not as much as
Kurucz's model.

The model we obtained from only the inversion of line profiles gives
rise to a continuum very similar (somewhat brighter) to that from
the theoretical model by Kurucz. This is a remarkable property, because
the MISS model was constructed from line profiles normalized to
the local continuum without considering any data on the absolute flux at any
wavelength or any other photometric information.

\subsubsection{Strong features}

The \ion{Ca}{1} $\lambda$ 6162.17 \AA\ line was already included in the
inversion procedure and its synthetic profile obtained using the MISS
model provides an excellent fit to the observed solar atlas, as can be
seen in the upper panel of Figure 7. This is also the case for the
other two \ion{Ca}{1} lines corresponding to the $4s4p$--$4s5s$
multiplet. It is also worthwhile noting the remarkably good fit
obtained for other lines surrounding the $\lambda$ 6162.17 \AA\ line,
not considered in the inversion, for which no necessarily
high-quality\footnote{The Vienna Atomic Line Database (VALD) provided
us with the transition probabilities, except for \ion{Ca}{1} 
$\lambda\lambda$ 6161 and 6162 \AA\ lines} transition probabilities have 
been used.

Another important comparison associated with strong features is shown
in the lower panel of Figure 7. The predicted wings of the sodium D
doublet, which are supposed to cover a wide depth range of the
photosphere and be formed under the LTE approximation (Covino et al.
1993), nicely reproduce the  observed solar profiles.

\subsubsection{Weak lines not included in the inversion}

Apart from the lines included in the inversion, it is of great interest
to compare predicted and observed profiles of other lines for which
accurate transition probabilities are available. A very suitable group
of lines for this purpose are the \ion{Fe}{1} lines measured by O'Brian
et al.  (1991) with an accuracy better than 5 \%, which belong to the
Meylan et al. (1993) selection, but are not included in the Oxford line
lists. Figure 8 shows this comparison for three \ion{Fe}{1} lines
located at $\lambda\lambda$ 4635.85, 6027.05, and 6481.87 \AA. We have
assumed an iron abundance of log N(Fe)$=$ 7.48, a macroturbulent
velocity of 1.7 km s$^{-1}$ (the values obtained from the inversion),
and a constant microturbulence of 0.6 km s$^{-1}$. The profiles were
computed in LTE in all cases. It can be seen that the best fits
correspond to MISS, followed very closely by VALC, and then by HSRA
(although these are NLTE models), Kurucz, and HOLMU, respectively.

It can be concluded then that the MISS model reproduces, in general,
several key observational constraints with a quality similar to that of
other well-known models and behaves better than several of these models
when focusing on particular comparisons.

\section{Application to other late-type stars}
\label{sec4}

As explained in the introduction, the final goal of this work is to
develop a spectroscopic tool for deriving semi-empirical model
photospheres for late-type stars in general, and for metal-poor stars
in particular. In principle, the information used here for the solar
case (normalized line profiles with very high resolution and high
signal-to-noise ratio) can be also available for other bright stars.
Allende Prieto et al. (1995) have shown preliminary observations of a
metal-poor star obtained with $R\equiv\lambda /\Delta\lambda\sim
170000$ and S/N$\sim$ 300--600. Observations of the same and other
stars with slightly higher resolution and better signal-to-noise ratios
have been already carried out and are being processed. These values of
$R$ and S/N are, however, significantly smaller than those associated
with the solar atlas we have used. Furthermore, the wavelength coverage
achieved in the stellar observations is much smaller than that
available for the Sun. It will be necessary then to test in detail what
happens in the inversion when using different numbers of lines combined
with spectra of lower quality. This is, in any case, a practical
problem which can be solved by using ultra-high resolution
spectrographs to observe very bright objects.

A different problem associated with this semi-empirical procedure is
related to the spectral synthesis itself in terms of knowledge of the
atomic and molecular opacities (especially for metal-poor stars),
effects of non-LTE, convection and overshooting, etc. A way for
partially minimizing these effects would be to use as many lines as
possible with accurate transition probabilities and damping constants,
and also to focus as much as possible on near-infrared lines where the
opacity problems decrease.

\section{Conclusions}
\label{sec5}

The Multiline Inversion of Stellar Spectra (MISS) procedure developed
here has been proved able to find an LTE temperature stratification
for the solar photosphere from normalized high-resolution line
profiles which reproduces the solar continuum and the
limb-darkening sufficiently well and is in excellent agreement with observed
strong and weak spectral lines with accurate line data. A better
comparison of its performance in reproducing the limb-darkening of the
wings of strong lines with accurate damping constants is now being studied.

It is found that none of the input abundances (Anders \& Grevesse 1989)
for chromium, calcium, and titanium need to be changed to find the best
fit to the observations, but the iron abundance is preferred to vary
from the initially assumed value (7.67, in the scale where the hydrogen
abundance is 12) to the lower (approximately meteoritic) abundance
7.48.

A better fit to the solar flux line spectrum would require the
abandonment of  the hypothesis of hydrostatic equilibrium  and the
introduction of multi-component models or velocity fields inducing
asymmetries in the line profiles. Our work in the near future will
consider this improvement.

This procedure of semi-empirical modeling of stellar photospheres can
potentially be applied to other late-type stars. While other classical
methods of empirical modeling would require observations not available
for stars (the limb-darkening cannot be measured yet, and flux
calibrated spectra are affected by a number of larger uncertainties),
the normalized line profiles used by MISS are already at hand.

\acknowledgements

We express our gratitude to L. R. Bellot Rubio, M. Collados, E.
Simmoneau, and  J. C. del Toro Iniesta for  interesting comments on the
draft. We also thank T.  Meylan, who kindly provided a digital version
of his line list. This article has been corrected for English and style
by Terry Mahoney (Research Division, IAC). NOS/Kitt Peak FTS data used
here were produced by NSF/NOAO. This research has made use of the VALD
database and was partially supported by the Spanish DGES under projects
PB95-1132-C02-01 and PB95-0028.

\clearpage

\appendix
\section{Evaluation of the rotation profile}

The effect of the rotation on the spectral profile has been estimated by taking
advantage of the knowledge of the specific intensity of the local continuum for
each spectral line $I_c(\lambda,\mu_i),\;\; i=1, \cdots, n$, where $n$ is the
number of points of the quadrature used.  Let $r_{\mu_i}=\sqrt{1-\mu_i^2},
\;\; i=1, \cdots, n$ be the normalized radius corresponding to the point with
cosine of the heliocentric angle $\mu_i$.  The stellar disk is considered 
divided into $n$  annuli with internal and external radii $r_i$ and $R_i$,
respectively, defined as follows: $r_1=0$, $r_i=({r_{\mu_{i-1}}+r_{\mu_i}})/2,
\;\; i=2, \cdots, n$, $R_i=r_{i+1}, \;\; i=1, \cdots, n-1$, and $R_n=1$. Let us
assume that  $I_c(\lambda,\mu)$ varies linearly between $\mu_i$ and
$\mu_{i+1}$, in the following way:

\begin{eqnarray}
I_c(\lambda,\mu) = a_i \; \mu + b_i,         \; &{\rm with}& \;\; 
\mu_{i-1} \geq \mu \geq \mu_{i+1}, \;\; i=2, \cdots, n-1 \nonumber \\
I_c(\lambda,\mu) = a_2 \; \mu + b_2,         \; &{\rm with}& \;\; 
\mu \geq \mu_{1}\\
I_c(\lambda,\mu) = a_{n-1} \; \mu + b_{n-1},  &{\rm with}& \;\; 
\mu \leq \mu_{n} \nonumber
\end{eqnarray}

The line-of-sight component of the rotation velocity will be the same for all
points located at an equal distance $x$ from the central meridian: $v_x=v
\sin{\hat\i} \; x$. The flux crossing the fraction of area of the $i$-th annulus
at a distance between $x$ and $x+\d x$ is given by
\begin{equation}
F_{x,i}=a_i \; \left[ y_i \; \mu_{r_{i}} - y_{i-1} \; \mu_{r_{i-1}} + 
(1-x^2) \; (\theta_i - \theta_{i-1}) \right] + 2 \; b_i (y_i - y_{i-1}),
\end{equation} 

\noindent for $i=1, \cdots, n$, where $\mu_{r_i}=\sqrt{1-{r_i}^2}$,
$\mu_{r_0}=0$,  $\theta_i= \arcsin ( \frac {y_i} {\sqrt{1-x^2} } )$, and
$y_i=\sqrt{r_i^2-x^2}$. From these expressions the rotation profile for each
annulus can be written as 
\begin{equation}
G_i= \frac{ F_{x,i} \; c}  { \lambda_0 \; v\sin \hat\i \; x \; \int_{r_i}^{R_i}
F_{x,i} \; \d x  }.
\end{equation}

\clearpage

\centerline{\bf FIGURE CAPTIONS}

\bigskip
%fig1
\figcaption[]{Three steps in the inversion procedure. From a
constant-temperature atmosphere (a, dotted lines) resulting in no
spectral lines, to the final step (c, solid lines) where the line
profiles are well reproduced.  Nine of the 40 lines employed for the
inversion are displayed; {\it upper panel} (left to right): Fe {\sc I}
$\lambda\lambda$ 6082, 6151, 6173, and 6200 \AA; {\it lower
panel}: Ca {\sc I} $\lambda\lambda$ 6166, 6455, and 6499 \AA, and Cr
{\sc I} $\lambda\lambda$ 4801 and 4964 \AA.}

%fig2
\figcaption[]{A detailed example of a synthetic line profile emerging
from the obtained  model atmosphere (solid line) compared to the solar
observations (circles). The reproduction of the Ca {\sc I} $\lambda$
6500 \AA\ observed profile is excellent. The goodness of the fit is limited
by the line asymmetries originating from convective motions.}

%fig3
\figcaption[]{Comparison between the solar photospheric temperature structure 
obtained using the inversion method (labeled MISS) and other 
classical solar models. MISS tends to be cooler in the outer layers and
hotter in the inner ones.}

%fig4
\figcaption[]{Comparison between the continuum limb-darkening observed
and predicted by different models in the optical spectral region.  The
model HSRA, empirically built from continuum observations,  follows the
observations very closely at all wavelengths while the model from the
inversion of line profiles reproduces the limb darkening acceptably.}

%fig5
\figcaption[]{The center-to-limb variations of the equivalent widths
(normalized to their value at the disk center) of three lines employed in the
flux inversion and measured by Balthasar (1988) are closer to the MISS
predictions than to any other model in this comparison.}

%fig6
\figcaption[]{Comparison between the solar flux measurements of Neckel
\& Labs (1984; filled triangles) and the pseudo-continuum estimated by
Kurucz et al.  (1984; dots), and the true continua predicted by
different models. The
 pseudo-continuum of Kurucz et al.  establishes a lower limit to the
 true continuum, which is well-satisfied by three of the models
considered:  MISS, Kurucz, and HOLMU.}

%fig7
\begin{figure}[ht]
\caption[]{{\it Upper panel}: the observed profile (solid line) of the
Ca {\sc I} $\lambda$ 6162 \AA\ line, whose wings (enhanced in the
graph) were included in the inversion, is nicely reproduced by the
resulting model. There is also a good agreement with the other
surrounding lines, which were not included in the inversion and without
accurate oscillator strengths. {\it Lower panel}: a very good agreement is
also found for the wings of the Na {\sc I} D lines (not included in
the inversion). For both the Na {\sc I} D  and Ca {\sc I}
$4s4p-4s5s$ lines,  accurate damping constants available from empirical
measurements and/or theoretical estimates exist in the literature and
have been used in the synthesis.}
\end{figure}

%fig8
\figcaption[]{The meteoritic iron abundance and the laboratory
transition probabilities measured by O'Brian et al. (1991) give an
independent test for the reliability of the MISS model. Displayed
are three neutral iron lines with clean profiles (following Meylan et
al. 1993), not included in the inversion line sample: $\lambda\lambda$
4635.85, 6027.05, and 6481.87 \AA. The observed profiles are marked as
dots while the other line styles are the same as in Fig. 3.}

\clearpage

%table 1
\begin{deluxetable}{cccccccc}
\tablecaption{Spectral lines used in the inversion. \label{table1}}
\tablehead{
\colhead{Element} & \colhead{Wavelength} & \colhead{Exc. Pot.} & 
\colhead{$\log gf$} &
\colhead{Element} & \colhead{Wavelength} & \colhead{Exc. Pot.} & 
\colhead{$\log gf$} \nl
 & \colhead{(\AA)} & \colhead{(eV)} &  &
 & \colhead{(\AA)} & \colhead{(eV)} &  }
\startdata 
\ion{Ca}{1} & 4578.557 & 2.52 & --0.697 &\ion{Ti}{1} & 5922.115 & 1.05 & --1.410\\
\ion{Ca}{1} & 5512.986 & 2.93 & --0.447 &\ion{Ti}{1} & 6092.799 & 1.89 & --1.323\\
\ion{Ca}{1} & 6161.297 & 2.52 & --1.266 &\ion{Ti}{1} & 6258.109 & 1.44 & --0.299\\
\ion{Ca}{1} & 6162.183 & 1.89  & --0.097 &\ion{Ti}{1} & 6303.762 & 1.44 & --1.510\\
\ion{Ca}{1} & 6166.441 & 2.52 & --1.142 &\ion{Ti}{1} & 6312.244 & 1.46 & --1.496\\
\ion{Ca}{1} & 6455.604 & 2.52 & --1.290 &\ion{Ti}{1} & 7357.735 & 1.44 & --1.066\\
\ion{Ca}{1} & 6499.656 & 2.52 & --0.818 &\ion{Fe}{1} & 4602.006 & 1.61 & --3.150\\
\ion{Cr}{1} & 4801.028 & 3.12 & --0.131 &\ion{Fe}{1} & 5225.533 & 0.11 & --4.790\\
\ion{Cr}{1} & 4964.931 & 0.94 & --2.527 &\ion{Fe}{1} & 5247.057 & 0.09 & --4.950\\
\ion{Cr}{1} & 5272.002 & 3.45 & --0.422 &\ion{Fe}{1} & 5916.254 & 2.45 & --2.990\\
\ion{Cr}{1} & 5300.751 & 0.98 & --2.129 &\ion{Fe}{1} & 5956.700 & 0.86 & --4.610\\
\ion{Cr}{1} & 5312.859 & 3.45 & --0.562 &\ion{Fe}{1} & 6082.715 & 2.22 & --3.570\\
\ion{Cr}{1} & 5787.922 & 3.32 & --0.083 &\ion{Fe}{1} & 6151.623 & 2.18 & --3.300\\
\ion{Cr}{1} & 7355.899 & 2.89 & --0.285 &\ion{Fe}{1} & 6173.342 & 2.22 & --2.880\\
\ion{Ti}{1} & 4758.122 & 2.25 & \phs 0.481 &\ion{Fe}{1} & 6200.321 & 2.61 & --2.440\\
\ion{Ti}{1} & 4759.274 & 2.25 &  \phs 0.570 &\ion{Fe}{1} & 6297.801 & 2.22 & --2.740\\
\ion{Ti}{1} & 5113.445 & 1.44 & --0.727 &\ion{Fe}{1} & 6481.878 & 2.28 & --2.980\\
\ion{Ti}{1} & 5295.781 & 1.05 & --1.577 &\ion{Fe}{1} & 6498.945 & 0.96 & --4.700\\
\ion{Ti}{1} & 5490.154 & 1.46 & --0.877 &\ion{Fe}{1} & 6750.161 & 2.42 & --2.620\\
\ion{Ti}{1} & 5866.457 & 1.07 & --0.784 &\ion{Fe}{1} & 6978.861 & 2.48 & --2.500\\
\enddata
\end{deluxetable}


\begin{references}


\reference{} Allende Prieto, C., \& Garc\'\i a L\'opez, R. J. 1998, \aaps\ (in
press)

\reference{viena} Allende Prieto, C., Garc\'{\i}a L\'opez, R. J.,
Lambert, D. L.,  \& Gustafsson, B.  1995, in IAU Symp. 176, Stellar
Surface Structure, Poster Proceedings, ed. K. G. Strassmeier (Vienna:
Institut f\"ur Astronomie der Universit\"at Wien), 107

\reference{pressure} Allende Prieto, C., Garc\'{\i}a L\'opez, R. J., \& Trujillo
Bueno, J. 1997, \apj, 483, 941

\reference{angre} Anders, E., Grevesse, N. 1989,  Geochimica et Cosmochimica
Acta, 53, 197

\reference{} Avrett, E. 1996, in IAU Symp. 176, Stellar Surface
Structure,  ed. K. G. Strassmeier \& J. L. Linsky (Dordrecht: Kluwer),
503

\reference{balthasar88} Balthasar, H. 1988, \aaps, 72, 473


\reference{bellot} Bellot Rubio L. R., Ruiz Cobo B., \& Collados M. 1997, \apj,
  478, L45

\reference{} Bellot Rubio L. R., Ruiz Cobo B., \& Collados M. 1998, \apj\
 (submitted)

\reference{blackwell} Blackwell, D. E., \& Shallis, M. J 1979,  \mnras, 186, 669

\reference{burlov} Burlov-Vasiljev, K. A. Gurtovenko, E. A., Matvejev, Yu. B.
1995, Solar Phys., 157, 51
  
\reference{}
Collados,  M., Mart\'\i nez Pillet,  V., Ruiz Cobo,  B., del Toro Iniesta,  J.C., \&
V\'azquez,  M. 1994, \aap,   291, 622

\reference{} Covino, E., Gomez, M. T., Severino, G., \& Franchini, M.  1993,
in  ASP Conf. Series, vol. 40, Inside the Stars, eds. W. W. Weiss \&  A. 
Baglin (San Francisco: ASP), 190

\reference{dravins81} Dravins, D., Lindegren, L., \& Nordlund, \AA.
1981, A\&A, 96, 345

\reference{} Dravins, D., \& Nordlund, \AA\ 1990a, \aap, 228, 184

\reference{} Dravins, D., \& Nordlund, \AA\ 1990b, \aap, 228, 203

\reference{hsra} Gingerich, O., Noyes, R. W., Kalkofen, W., \& Cuny, Y., 1971,
Solar Phys., 18, 347

\reference{gray80} Gray, D. F. 1980, ApJ, 235, 508

\reference{} Gustafsson, B., \& J\o rgensen, U. G. 1994, \aapr, 6, 19

\reference{holmu74} Holweger, H.,  \& M\"uller, E. A. 1974, Solar Phys.,
39, 19

\reference{kurucz92} Kurucz, R. L. 1992,  in  IAU Symp. 149, The
Stellar Populations of Galaxies, ed. B. Barbuy \& A. Renzini
(Dordrecht: Kluwer),  225

\reference{kuruczisot93} Kurucz,  R. L.   1993, Physica Scripta, T47,
110

\reference{kuruczatlas84} Kurucz, R. L., Furenlid, I., Brault, J., \&
Testerman, L. 1984, NOAO Atlas No. 1, The Solar Flux Atlas from 296 to
1300 nm (Sunspot, NM: National Solar Observatory)

\reference{ } Marquardt, D. W. 1963, J. Soc. Ind. Appl. Math.,  11, 431

\reference{} M\"ackle, R., Holweger, H, Griffin, R., \& Griffin, R. 1975, \aap,
38, 239 

\reference{meylan} Meylan, T., Furenlid, I., Wiggs, M. S., Kurucz, R. L. 1993, 
\apjs, 85, 163

\reference{neckel84} Neckel, H., \& Labs, D. 1984, Solar Phys., 90, 205

\reference{neckel94} Neckel, H., \& Labs, D. 1994, Solar Phys., 153, 91

\reference{obrian} O'Brian, T. R., Wickliffe, M. E., Lawler, J. E., Whaling, W.,
\& Brault, J. W. 1991, J. Opt. Soc. Am. B, 8,  1185

\reference{} Press, W. H., Flannery, B. P., Teukolsky, S. A., \& Vetterling, W. T.
1986, Numerical Recipes (Cambridge: Cambridge University Press)

\reference{} Rast, M. P., Nordlund, \AA, Stein, R., \& Toomre, J. 1993, \apjl,
408, L53

\reference{}Rodr\'\i guez Hidalgo, I., Ruiz Cobo, B., \& Collados, M. 1996, 
in ASP Conf. Ser. 109, Cool Stars, Stellar Systems, and the Sun, 9$^{\rm th}$ 
Cambridge Workshop, ed. R. Pallavicini, 
\& A. K. Dupree (San Francisco: ASP),  155

\reference{} Rodr\'\i guez Hidalgo, I., Ruiz Cobo, B., \& Collados, M. 1997,
Solar Phys., 172, 72

\reference{} Ruiz Cobo, B., Rodr\'\i guez Hidalgo, I., \& Collados, M. 1997,
\apj, 488,  462

\reference{} Ruiz Cobo, B., \& del Toro Iniesta, J. C. 1992, \apj, 398, 375

\reference{} Ruiz Cobo, B., \& del Toro Iniesta, J. C. 1994, \aap, 283, 129

\reference{} Ruiz Cobo, B., del Toro Iniesta, J. C., Rodr\'\i guez Hidalgo, I.,
Collados, M., \& S\'anchez Almeida, J. 1995, JOSO Annual Report, 162


\reference{} Ruiz Cobo, B., del Toro Iniesta, J. C., Rodr\'\i guez
Hidalgo, I., Collados, M., \& S\'anchez Almeida, J. 1996, in ASP Conf.
Ser. 109, Cool Stars, Stellar Systems, and the Sun,  9$^{\rm th}$
Cambridge Workshop ASP Conference, eds. R. Pallavicini \& A.K. Dupree
(San Francisco: ASP),  155

\reference{} Ruland, F., Holweger, H., Griffin, R., Griffin, R., \& Biehl, D.
1980, \aap, 92, 70

\reference{} Socas-Navarro, H., Trujillo Bueno, J., Ruiz Cobo, B., \&
Shchukina, N. G. 1996, JOSO Annual Report, 86

\reference{}
Socas-Navarro, H., Ruiz Cobo, B., \&  Trujillo Bueno, J. 1998, \apj\  (in
preparation)

\reference{}
Spielfiedel, A., Feautrier, N., Chambaud, G., \& L\'evy, B. 1991, J. Phys. B, 24, 4711

\reference{}
Takeda, Y. 1995, \pasp, 47, 337

\reference{} 
del Toro Iniesta, J. C., \& Ruiz Cobo, B. 1995, in La polarim\'etrie, outil pour
l'\'etude de l'activit\'e magn\'etique solaire et stellaire, eds. N. Mein \& S. 
Sahal-Br\'echot (Paris: Observatoire de Paris-Meudon), 127

\reference{} 
del Toro Iniesta, J. C., \& Ruiz Cobo, B. 1996, Solar Phys., 164, 169

\reference{} del Toro Iniesta, J.C., \& Ruiz Cobo, B. 1997, in Forum
THEMIS, Observatoire de Paris-Meudon, eds. N. Mein \& S.
Sahal-Br\'echot (Paris: Observatoire de Paris-Meudon),  93

\reference{} 
del Toro Iniesta, J. C., Tarbell T. D., \& Ruiz Cobo, B. 1994, \apj, 436, 400

\reference{valc} Vernazza, J. E., Avrett, E. H., \& Loeser, R. 1981, \apjs, 45, 635

\reference{}
Westendorp Plaza,  C., del Toro Iniesta, J. C., Ruiz Cobo,  B., Mart\'\i nez
Pillet,  V., Lites,  B. W., \& Skumanich, A. 1997a, in 1$^{\rm st}$ Advances
in Solar Physics Euroconference. Advances in the Physics of Sunspots,  eds. 
B. Schmieder, J. C. del Toro Iniesta, \& M. V\'azquez, ASP Conf. Ser.
Vol. 118, 197

\reference{}
Westendorp Plaza, C., del Toro Iniesta J. C., Ruiz Cobo, B., Mart\'\i nez
Pillet, V., Lites, B. W., \& Skumanich, A. 1997b, in 1$^{\rm st}$ Advances
in Solar Physics Euroconference. Advances in the Physics of Sunspots,  eds. 
B. Schmieder, J.C. del Toro Iniesta, \& M. V\'azquez, ASP Conf. Ser.
Vol. 118, 202

\reference{}
Westendorp Plaza, C., del Toro Iniesta, J. C., Ruiz Cobo, B., 
Mart\'\i nez Pillet, V., Lites, B. W., \& Skumanich, A. 1997c, \nat, 389, 47
 
\reference{}
Westendorp Plaza, C., del Toro Iniesta J. C., Ruiz Cobo, B., 
Mart\'\i nez Pillet, V., Lites B. W., \& Skumanich, A. 1998, \apj, 494 (in press)

\reference{}
Wood K. \& Fox G.K. 1995, Inverse Problems, 11, 795


\end{references}
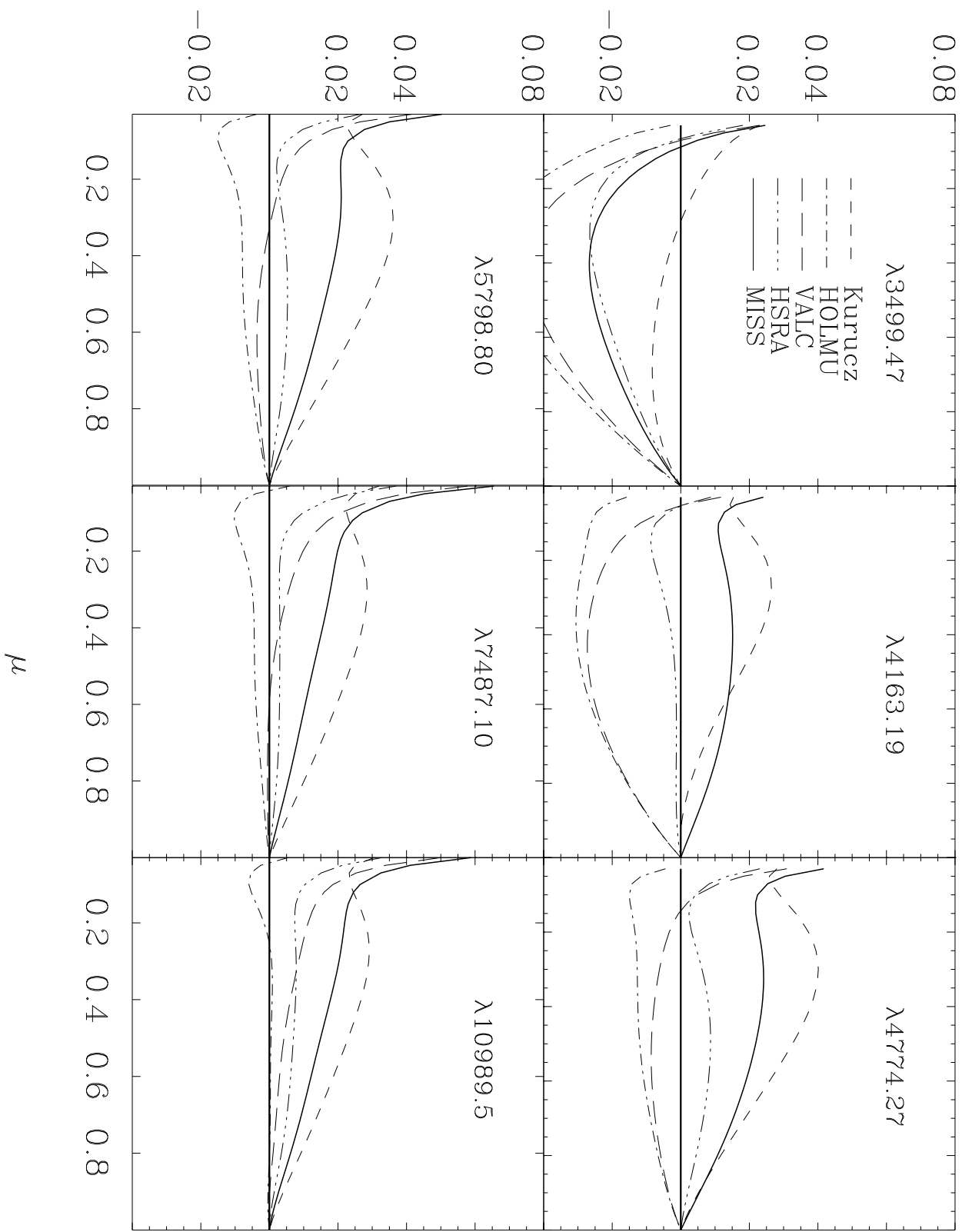
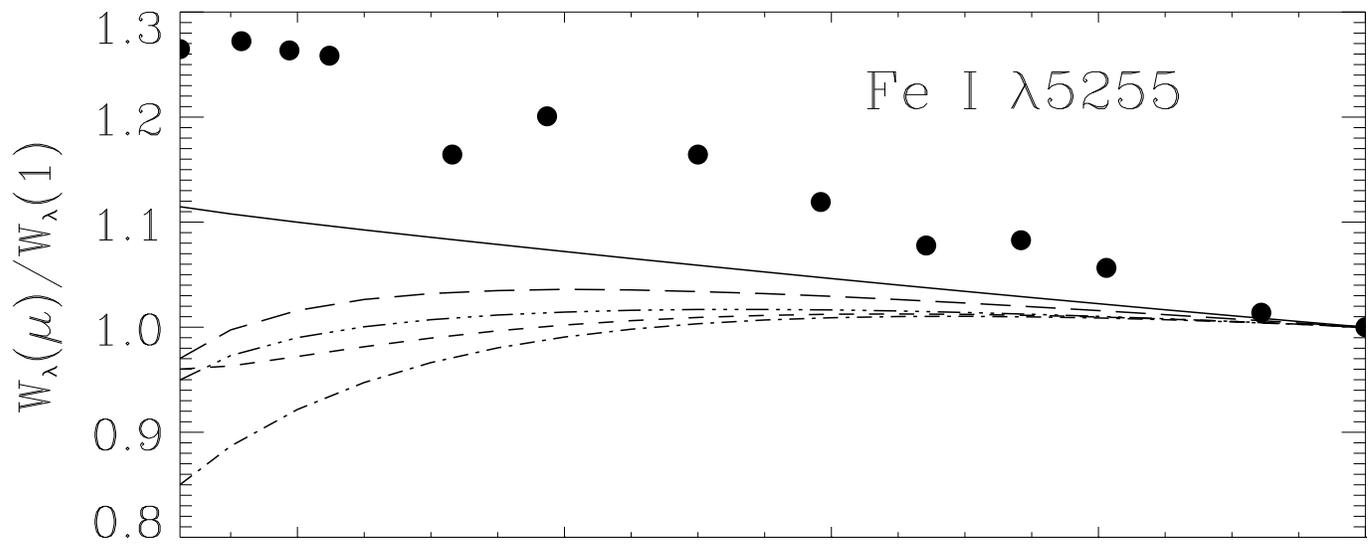
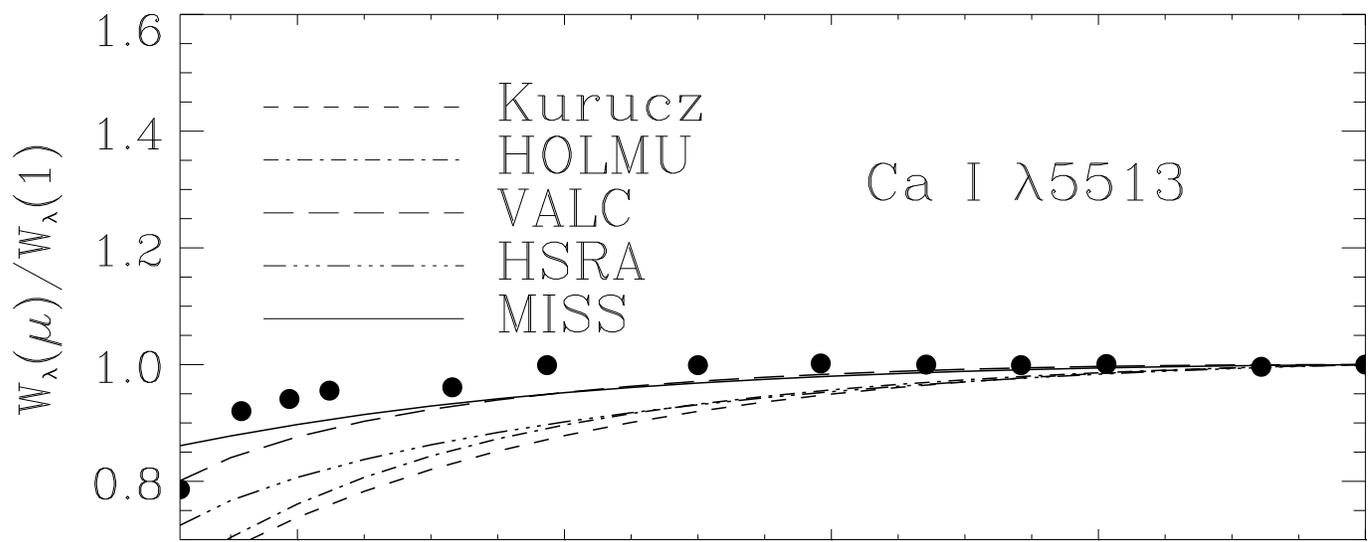
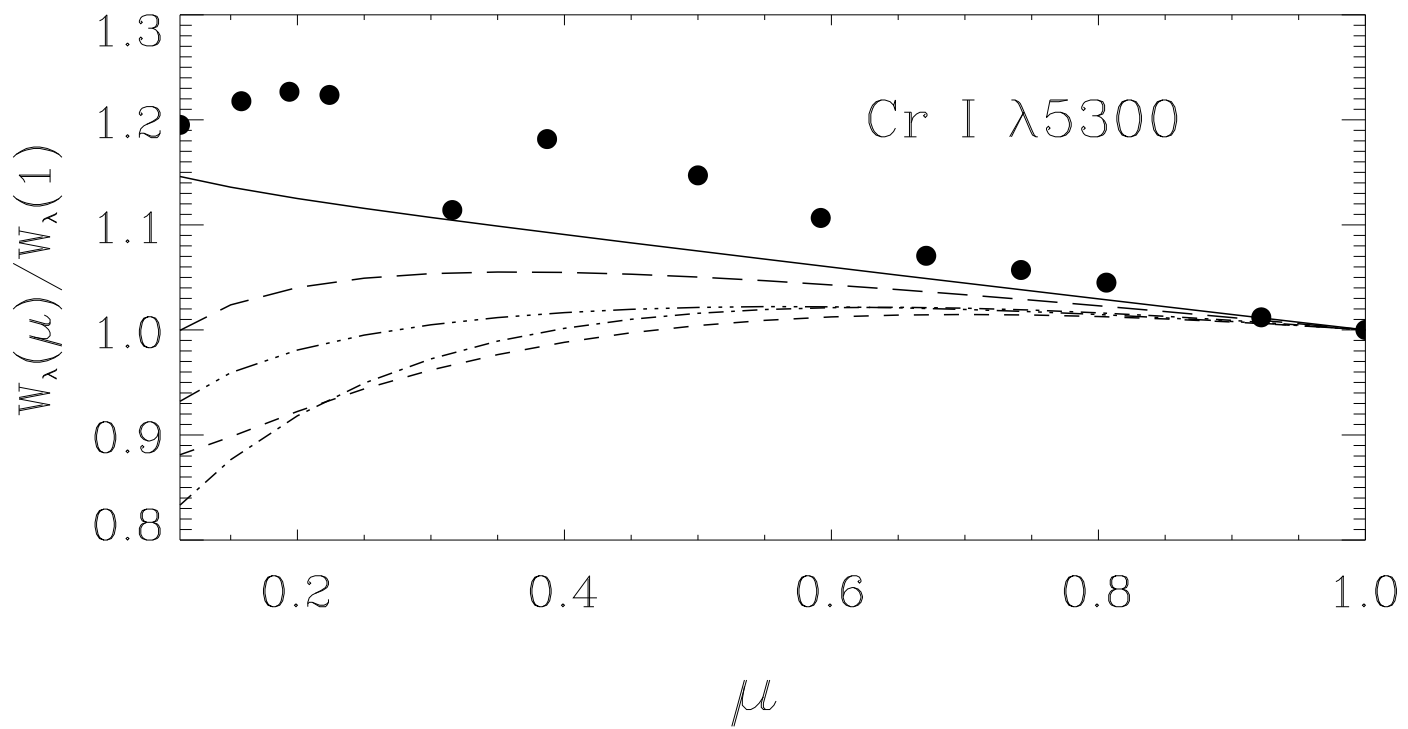
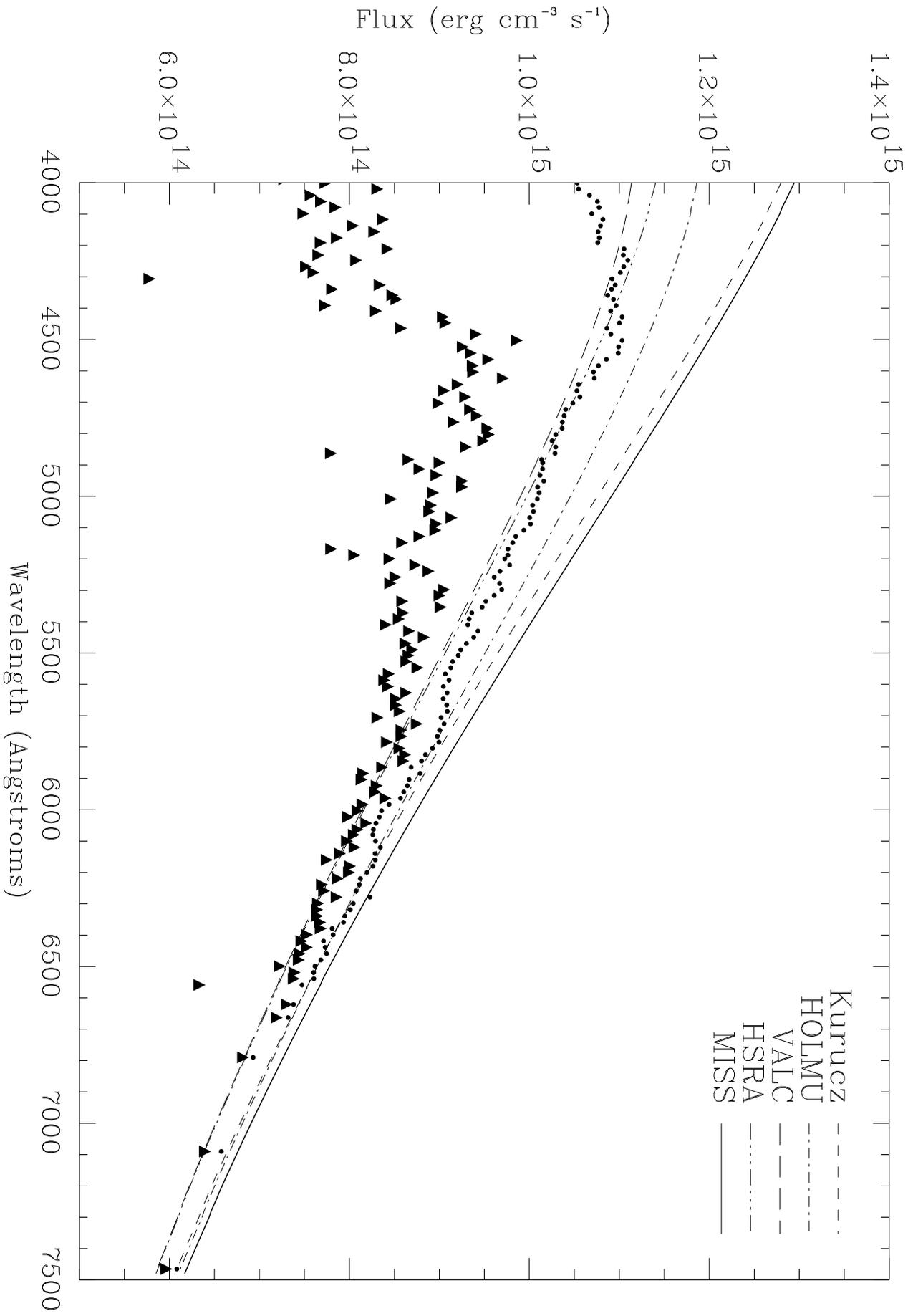
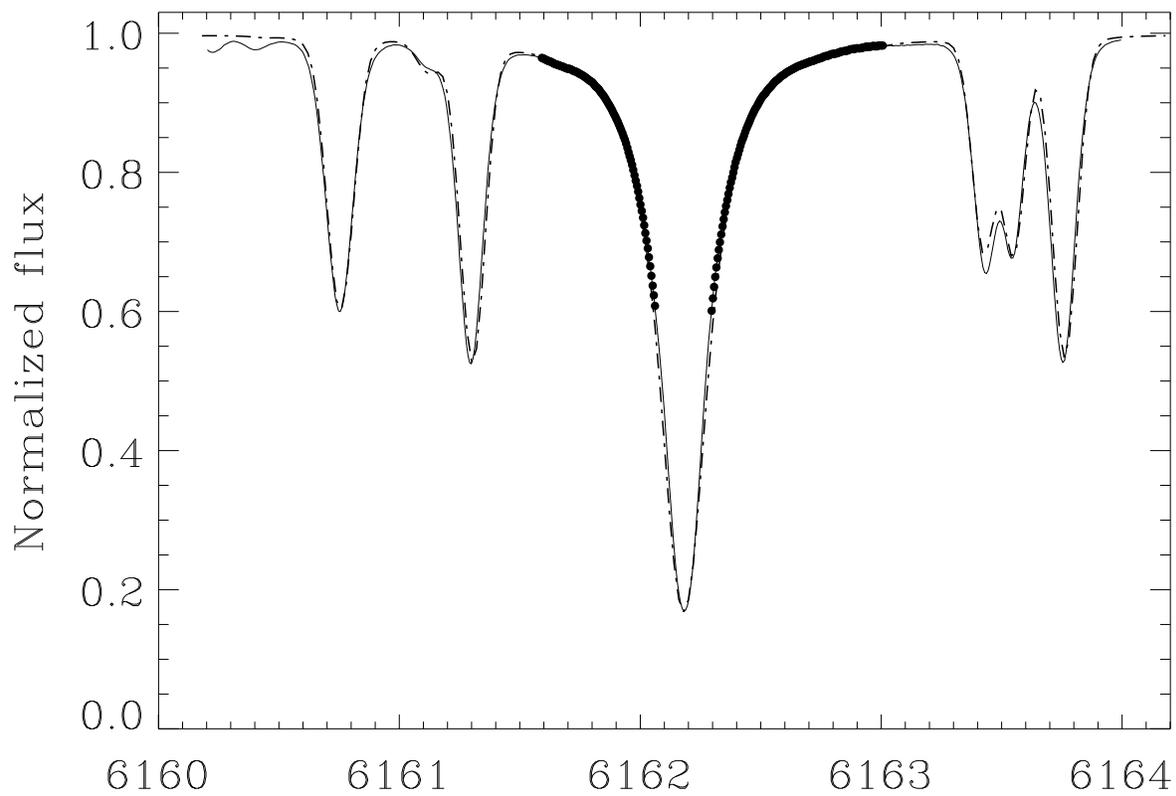
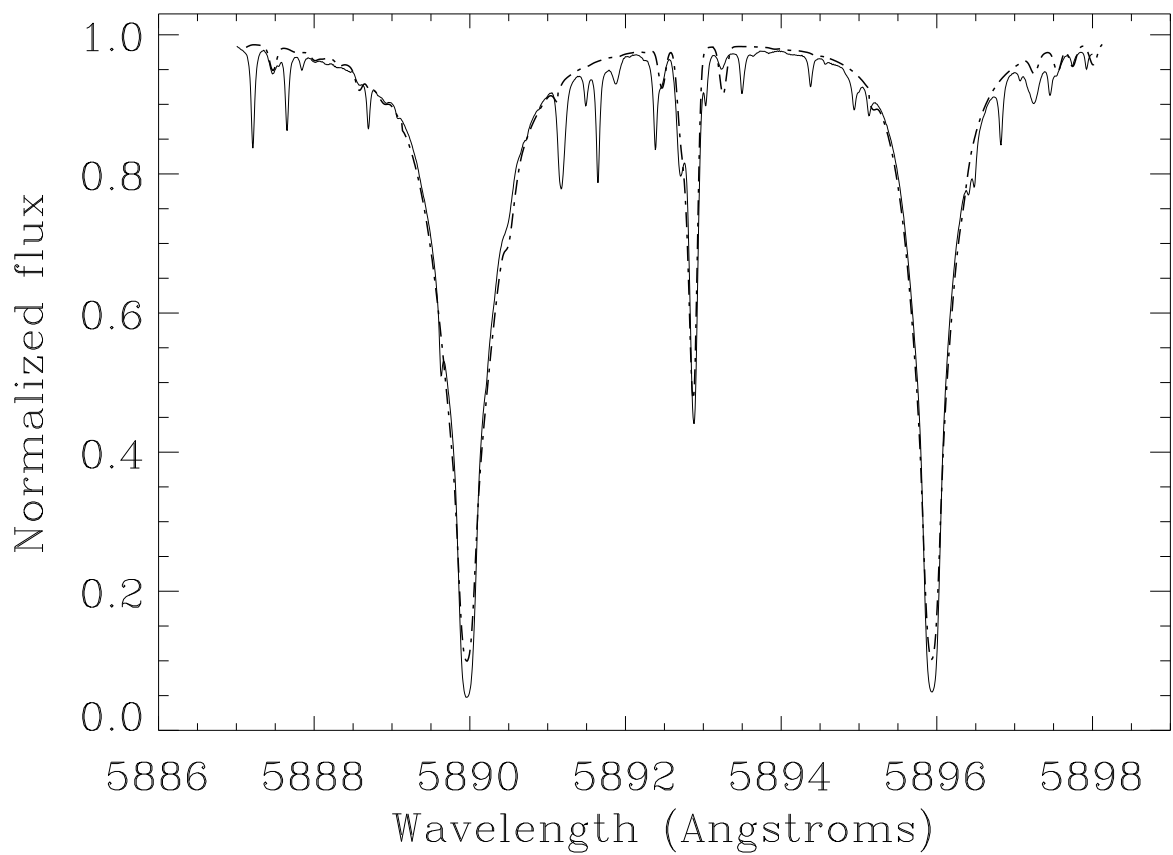
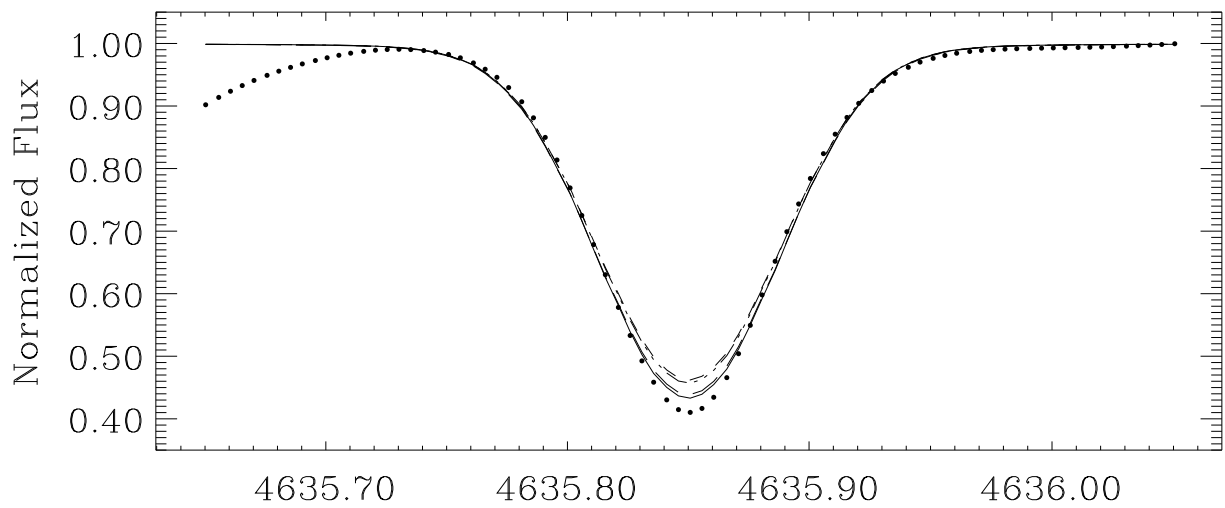
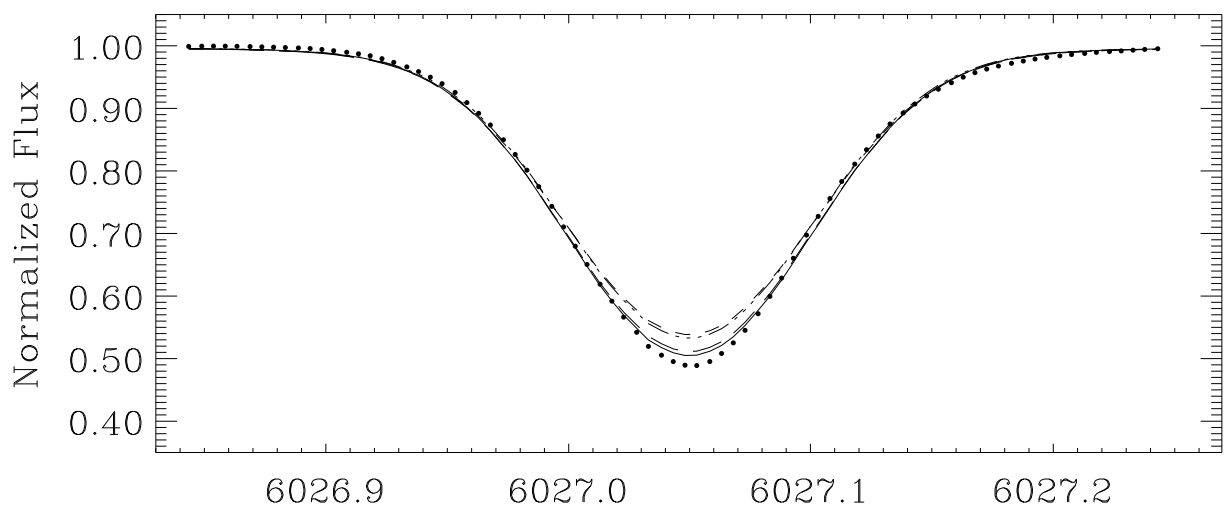
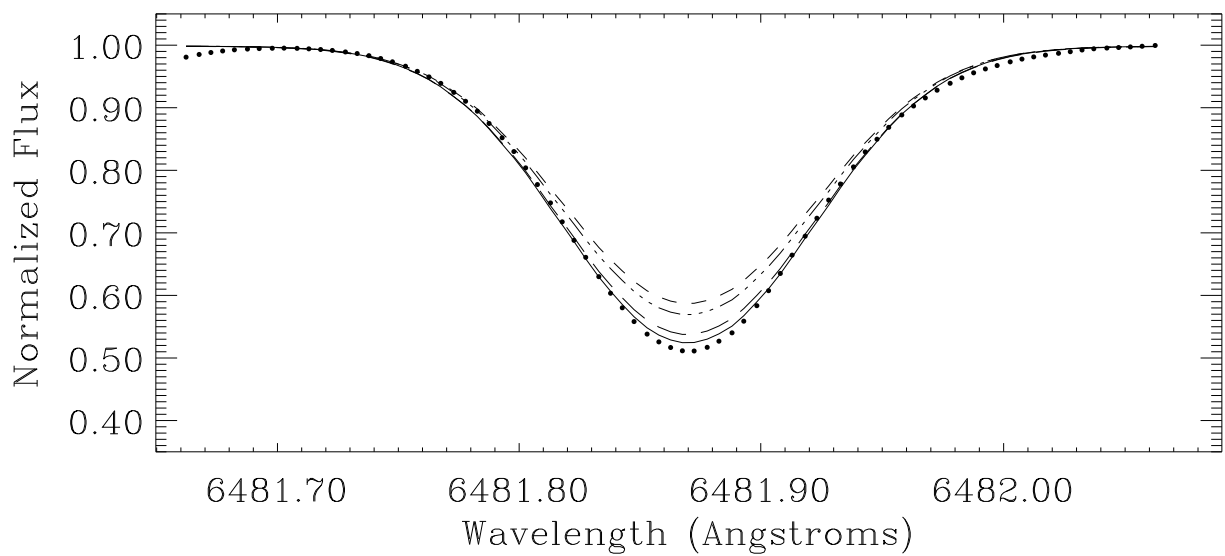
\end{document}